\documentclass[twocolumn]{aastex62}

\shorttitle{Magnetic fields of new NS}
\shortauthors{Soker, N.}
\def \s{~\rm{s}}
\def \km{~\rm{km}}

\def \G{~\rm{G}}

\def \erg{~\rm{erg}}

\begin{document}

\title{Amplifying magnetic fields of a newly born neutron star by stochastic angular momentum accretion in core collapse supernovae}

\correspondingauthor{Noam Soker}
\email{soker@physics.technion.ac.il}

\author[0000-0003-0375-8987]{Noam Soker}
%\affil{Departmeמt of Physics, Technion, Haifa 3200003, Israel}
\affiliation{Departmeמt of Physics, Technion, Haifa, 3200003, Israel}
\affiliation{Guangdong Technion Israel Institute of Technology, Shantou 515069, Guangdong Province, China}
%\collaboration{(AAS Journals Data Scientists collaboration)}

%% Note that the \and command from previous versions of AASTeX is now
%% depreciated in this version as it is no longer necessary. AASTeX
%% automatically takes care of all commas and "and"s between authors names.

%% AASTeX 6.2 has the new \collaboration and \nocollaboration commands to
%% provide the collaboration status of a group of authors. These commands
%% can be used either before or after the list of corresponding authors. The
%% argument for \collaboration is the collaboration identifier. Authors are
%% encouraged to surround collaboration identifiers with ()s. The
%% \nocollaboration command takes no argument and exists to indicate that
%% the nearby authors are not part of surrounding collaborations.

%% Mark off the abstract in the ``abstract'' environment.
\begin{abstract}
I present a novel mechanism to boost magnetic field amplification of newly born neutron stars in core collapse supernovae. In this mechanism, that operates in the jittering jets explosion mechanism and comes on top of the regular magnetic field amplification by turbulence, the accretion of stochastic angular momentum in core collapse supernovae forms a neutron star with strong initial magnetic fields but with a slow rotation.
The varying angular momentum of the accreted gas, which is unique to the jittering jets explosion mechanism, exerts a varying azimuthal shear on the magnetic fields of the accreted mass near the surface of the neutron star. This, I argue, can form an amplifying effect which I term the stochastic omega (S$\omega$) effect. In the common $\alpha \omega$ dynamo the rotation has constant direction and value, and hence supplies a constant azimuthal shear, while the convection has a stochastic behavior. In the S$\omega$ dynamo the stochastic angular momentum is different from turbulence in that it operates on a large scale, and it is different from a regular rotational shear in being stochastic. {{{{ The basic assumption is that because of the varying direction of the angular momentum axis from one accretion episode to the next, the rotational flow of an accretion episode stretches the magnetic fields that were amplified in the previous episode. I estimate the  amplification factor of the S$\omega$ dynamo alone to be $\approx 10$.  }}}}
I speculate that the S$\omega$ effect accounts for a recent finding that many neutron stars are born with strong magnetic fields.
\end{abstract}

%% Keywords should appear after the \end{abstract} command.
%% See the online documentation for the full list of available subject
%% keywords and the rules for their use.
\keywords{stars: neutron --- stars: magnetic field --- (stars:) supernovae: general}

%% From the front matter, we move on to the body of the paper.
%% Sections are demarcated by \section and \subsection, respectively.
%% Observe the use of the LaTeX \label
%% command after the \subsection to give a symbolic KEY to the
%% subsection for cross-referencing in a \ref command.
%% You can use LaTeX's \ref and \label commands to keep track of
%% cross-references to sections, equations, tables, and figures.
%% That way, if you change the order of any elements, LaTeX will
%% automatically renumber them.
%%
%% We recommend that authors also use the natbib \citep
%% and \citet commands to identify citations.  The citations are
%% tied to the reference list via symbolic KEYs. The KEY corresponds
%% to the KEY in the \bibitem in the reference list below.

% ==========================================================
\section{Introduction}
\label{sec:intro}
% ==========================================================

Core collapse supernovae (CCSNe) occur when the core of massive stars collapses to form a neutron star (NS) or a black hole (e.g., \citealt{WoosleyWeaver1986}). A fraction of the gravitational energy that the collapsing gas releases powers the explosion by ejecting the rest of the core and the envelope (e.g., \citealt{Janke2012}). There is no consensus yet on the processes that channel the gravitational energy to explosion, {{{{ as I discuss next. }}}}

There are two contesting theoretical models to channel the gravitational energy to explosion, the delayed neutrino mechanism \citep{BetheWilson1985} and the jittering jets explosion mechanism \citep{Soker2010}. There are, what I see as, some difficulties in the delayed neutrino mechanism (e.g., \citealt{Papishetal2015, Kushnir2015b}), {{{{ e.g., in the classical one-dimensional delayed neutrino mechanism the heating by neutrinos has no time to accelerate the ejecta to high energies \citep{Papishetal2015}. Three-dimensional effects seem to partially solve this problem (e.g., \citealt{Mulleretal2019Jittering}). }}}} There are also seemingly contradicting results in obtaining explosions with the desired explosion energy {{{{ as some obtain explosions (e.g., \citealt{Mulleretal2017, Vartanyanetal2019}) and some  do not (e.g. \citealt{OConnorCouch2018}). }}}} For these, I consider the jittering jets explosion mechanism as a more successful explosion model, {{{{ or the two mechanisms of heating by neutrinos and jittering jets should act together \citep{Soker2019JitSim}. }}}}

The jittering jets explosion mechanism includes a negative feedback component (e.g., \citealt{Gilkisetal2016, Soker2017RAA}), the jet feedback mechanism (for a review see \citealt{Soker2016Rev}), in the sense that when the jets expel the rest of the core and the envelope, they shut themselves down. {{{{ This implies that the explosion energy is of the order of, or several times, the binding energy of the core for an efficient feedback process, or many tens times the binding energy of the core in rare cases when the feedback process is not efficient \citep{Gilkisetal2016}. This expected behavior of the jittering jets explosion mechanism is compatible with observations.  }}}} 
  
As well, neutrino heating does play a role in the jittering jets explosion mechanism \citep{Soker2018KeyRoleB, Soker2019SASI, Soker2019JitSim}, but not the dominant role. When the pre-collapse core is slowly rotating the angular momentum of the accretion flow onto the newly born NS will be highly stochastic due to fluctuations in the convective regions of the pre-collapse core or envelope \citep{GilkisSoker2014, GilkisSoker2015, Quataertetal2019}, that the spiral standing accretion shock instability (SASI) modes (for studies of the spiral SASI see, e.g., \citealt{BlondinMezzacappa2007, Rantsiouetal2011, Iwakamietal2014, Kurodaetal2014, Fernandez2015, Kazeronietal2017}) further amplify. When the pre-collapse core is rapidly rotating, {{{{ (i.e., the specific angular momentum of the gas allows it to form an accretion disk around the newly born NS), }}}} the jittering will have relative small amplitudes around a fixed angular momentum axis \citep{Soker2017RAA}.

Many studies have found indications, like polarization in some CCSNe and the morphology of some supernova remnants, for some roles that jets play in possibly most CCSNe (e.g., \citealt{Wangetal2001, Maundetal2007, Smithetal2012, Lopezetal2011, Milisavljevic2013, Gonzalezetal2014, Marguttietal2014, Inserraetal2016, Mauerhanetal2017, GrichenerSoker2017, Bearetal2017, Garciaetal2017, LopezFesen2018}). As well, there are many studies of jet-driven CCSNe that do not consider jittering, and hence are aiming at rare cases, e.g., of progenitors having a very rapidly rotating pre-collapse core (e.g., \citealt{Khokhlovetal1999, Aloyetal2000, Hoflich2001, MacFadyen2001, Obergaulingeretal2006, Burrows2007, Nagakuraetal2011, TakiwakiKotake2011, Lazzati2012, Maedaetal2012, LopezCamaraetal2013, Mostaetal2014, Itoetal2015, BrombergTchekhovskoy2016, LopezCamaraetal2016, Nishimuraetal2017, Fengetal2018, Gilkis2018, Obergaulingeretal2018}).

The jittering-jets explosion mechanism differs from the processes that these studies of jets consider in having some unique properties.  (1) The jittering jets explosion mechanism supposes to explode all CCSNe with kinetic energies of $\ga 2 \times 10^{50}\erg$, and many (or even all) of CCSNe below that energy, rather than only a small percentage of all CCSNe (e.g., \citealt{Soker2016Rev}). (2) The pre-collapse core can have any value of rotation, from non-rotating to rapidly rotating, rather than having rapid rotation only (e.g., \citealt{GilkisSoker2014}). (3) The jets can be intermittent, and for slowly rotating pre-collapse cores they are also strongly jittering (i.e., having large variable directions; e.g., \citealt{Soker2017RAA}). (4)  The jets operate in a negative feedback mechanism. Namely, the jets reduce the accretion rate and hence their power while removing mass from the core and envelope (e.g., \citealt{Soker2016Rev}). (5) In cases of a strong jittering, each jet-launching episode is active for a short time and the direction of the jets changes rapidly. Therefore, the jets do not break out from the ejecta of the explosion (e.g., \citealt{PapishSoker2011}). In some cases of strong jittering the jets from the last jet-launching episode or two might break out of the ejecta and inflate two opposite small lobes (called ears) on the outskirts of the supernova remnant (e.g., \citealt{Bearetal2017, GrichenerSoker2017}).

In the present study I continue my exploration of the jittering jets explosion mechanism and I raise the possibility that in the jittering jets explosion mechanism there is a process that contributes to magnetic field amplification in the material that the newly born neutron star accretes. I term this the stochastic-$\omega$ (S$\omega$) effect.
I describe this effect in section \ref{sec:parameters}, and discuss some plausible typical quantitative parameters in section \ref{sec:StochasticOmega}.
{{{{ The S$\omega$ that I propose differs from the $\alpha \omega$ dynamo, as I explain in the following sections. In the $\alpha \omega$ dynamo the $\omega$ refers to the stretching of poloidal magnetic field lines to azimuthal lines by an ordered differential rotation, while the $\alpha$ effect refers to stochastic motion, like turbulence, that entangles the azimuthal magnetic fields to form poloidal magnetic field lines to close the dynamo cycle. }}}}

I mention that simulations that do not consider the jittering jets explosion mechanism also find the accretion of stochastic angular momentum onto the newly born NS (e.g.,  \citealt{Kazeronietal2016, Mulleretal2017}). Hence, the S$\omega$ effect might take place also in the neutrino driven explosion mechanism. However in the jittering jets explosion mechanism the S$\omega$ effect is a generic outcome. As well, the amplification of magnetic fields by the S$\omega$ effect might help the launching of jets. In the present study I scale quantities according to the expactation of the jittering jets explosion mechanism.

In section \ref{sec:Summary} I summarise the main results and discuss the general picture of forming NSs with strong magnetic fields and the broader relation to the jittering jets explosion mechanism.

%%% Some studies (e.g., \citealt{Masadaetal2015, Mostaetal2015, ObergaulingerAloy2017, Obergaulingeretal2018}) have taken the first direction in exploring the role of magnetic fields by high resolution simulations.

% ==========================================================
\section{The Stochastic-Omega (S$\omega$) effect}
\label{sec:parameters}
% ==========================================================

% ===============================
\subsection{General description}
\label{subsec:GeneralDescription}
% ===============================

I consider the following general flow of a CCSN where the pre-collapse core rotational velocity is low, and so the collapsing core gas that feeds the newly born NS has a stochastic angular momentum. The total mass that flows on to the NS during this phase is $\simeq 0.1-0.5 M_\odot$ (e.g., \citealt{PapishSoker2011}). The in-flowing gas has an initial stochastic angular momentum and magnetic fields from the pre-collapse core, that are further amplified in the unstable region behind the stalled shock (see section \ref{sec:intro} for relevant references).

Even that the stochastic specific angular momentum is less than the value required for a Keplerian velocity on the surface of the  NS, the flow behind the stalled shock amplifies magnetic fields via the regular mechanism of the $\alpha \omega$ dynamo (e.g., \citealt{Soker2018KeyRoleB, Soker2019SASI}). The two ingredients of the $\alpha \omega$ dynamo are the turbulent motion {{{{ that entangles the azimuthal magnetic fields to form poloidal magnetic field lines }}}}  (the $\alpha$ effect), and the differential rotation of the toroidal (azimuthal) flow {{{{ that stretches poloidal magnetic field lines to azimuthal lines. }}}} In the regular $\alpha \omega$ dynamo the direction of the angular momentum of the toroidal flow does not change. As well, the azimuthal velocity depends only the poloidal location $(\varpi,z)$, where $\varpi$ is the distance from the symmetry axis and $z$ is the distance from the equatorial plane.

Here I consider the case where the angular momentum axis changes in a stochastic manner, and I study the effect that this might have on magnetic field amplification.
I term this the Stochastic-$\omega$ (S$\omega$) effect.
Both the $\alpha$ effect and the $\omega$ effect still exist, and I argue below that the S$\omega$ effect adds to the magnetic field amplification during the periods when the angular momentum axis changes its direction.
According to the jittering jets explosion mechanism there are about ten to few tens of accretion episodes, and the variations in the angular momentum axis take place in short periods between consecutive accretion episodes (see relevant references in section \ref{sec:intro}).

\begin{figure} [ht!]
%\plotone{NSmagneticFig1.pdf}
\vskip -1.1 cm
\hskip -1.9 cm
%\begin{tabular}{cc}
{\includegraphics[scale=0.54]{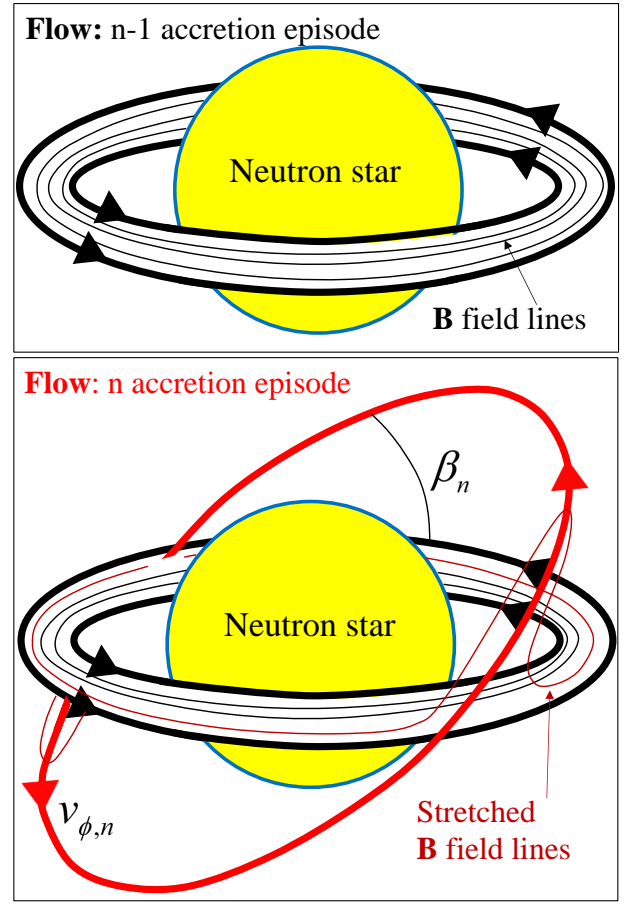}}
%\end{tabular}
\vskip -1.7 cm
\caption{A schematic drawing of the flow interaction between accretion episodes $n$ and $n-1$. The upper panel shows the flow in the accretion episode $n-1$ (thick black lines with arrows) and the magnetic field lines that this flows amplified as thin black lines. In the lower panel the thick red line schematically represents the flow of the inner part (early time) of the next accretion episode $n$. The angular momentum axes of the  two episodes are inclined to each other by an angle $\beta_n$. The field lines from episode $n-1$ that are in the interface of the two episodes are represented by a coloured line that is stretched by the flow of the $n$ episode.  In both episodes the toroidal flow is much thicker than what is drawn here, both in the radial direction away from the symmetry axis and perpendicular to the rotational plane. }
\label{fig:NSmagneticFig1}
\end{figure}

Specifically, in  the present study I focus on the amplification near the surface of the newly born NS just as these magnetic fields are dragged onto the NS. In Fig \ref{fig:NSmagneticFig1} I present a schematic description of the flow, showing the NS and two consecutive accretion episodes, number $n-1$ and $n$. The upper panel presents only the early accretion episode, where the differential rotation amplifies an azimuthal magnetic field that I present as thin magnetic field lines. This is the regular $\omega$ effect of the $\alpha \omega$ dynamo. In the lower panel I present one stream line (thick red line) of the next accretion flow, which has its angular momentum axis inclined by an angle $\beta_n$ to that of the flow in the  previous accretion episode. The new flow drags and stretches the magnetic field lines on the outer part of the flow of the previous accretion episode, which I represent by one thin red line. By that stretching the flow further amplifies the magnetic field. This is the S$\omega$ effect. The stochastic accretion of angular momentum of the S$\omega$ effect implies that the NS is born with slow rotation, but that nonetheless might have a strong magnetic field. Namely, it might be born as a slowly rotating magnetar.
% FFFFFFFFFFFFFFFFFFFFFFFFFFFFFFFFFFFFFFFFFFFFFFFFFFFFFFFFFFFF
% FFFFFFFFFFFFFFFFFFFFFFFFFFFFFFFFFFFFFFFFFFFFFFFFFFFFFFFFFFFF

% ===============================
\subsection{Relevant equations}
\label{subsec:Equations}
% ===============================

Consider the $\omega$ effect in the induction equation (e.g., \citealt{Priest1987})
\begin{equation}
\frac{\partial {\bf B}}{\partial t} = {\rm curl}({\bf v} \times {\bf B})+\eta \nabla^2  {\bf B},
\label{eq:Induction1}
\end{equation}
where ${\bf v}$ is the plasma velocity and $\eta$ is its magnetic diffusivity.
For a case where the main flow during the $n$ accretion episode is azimuthal $v_\phi$, i.e., toroidal flow along coordinate $\phi$, and neglecting magnetic dissipation, i.e., a very small value of $\eta$, the induction equation for the toroidal magnetic field component reads (e.g., \citealt{Priest1987})
\begin{equation}
\frac{\partial {B_\phi}}{\partial t} = 
R  {\bf B_p} \cdot {\bf \nabla}  \left( \frac{v_\phi}{R} \right),
\label{eq:Induction2}
\end{equation}
{{{{ where ${\bf B_p}$ is the poloidal component of the magnetic field, and $R$ is the distance from the rotation axis. What matters here is only the magnitude of equation (\ref{eq:Induction2}). In the above equation I assume that within each accretion episode the flow is steady and axisymmetric. However, in the  jittering jets explosion mechanism the flow is not steady and the axisymmetry axis changes direction between consecutive accretion episodes. }}}}

I assume that in accretion episode $n-1$ the differential rotation forms a strong azimuthal (toroidal) field, ${\bf B}_{n-1} \simeq B_{\phi1,n-1}$, where here $\phi 1$ is the direction of the toroidal flow of accretion episode $n-1$.
{{{{ In the jittering jets explosion mechanism there are about ten to few tens of accretion episodes (e.g., \citealt{PapishSoker2011}). }}}}
I take the azimuthal field of accretion episode $n-1$ to be the seed field of accretion episode $n$, where $n=2, . . .{\rm few}\times 10$. Let $\beta_n$ be the angle between the angular momentum axis of episode $n$ and episode $n-1$, such that the seed magnetic field at the beginning of accretion episode $n$ in the region where the two azimuthal consecutive flows cross each other is
\begin{equation}
{\bf B}_{n, 0} = B_{\phi1,n-1} \cos \beta_n \hat \phi
+B_{\phi1,n-1} \sin \beta_n \hat z
\label{eq:seedN}
\end{equation}
where here $\phi$ is the direction of the toroidal flow in episode $n$, and $\hat z$ is a unit vector perpendicular to the toroidal plane.

For the gradient of the velocity in the relevant direction $\hat z$ I take
\begin{equation}
{\bf \nabla} \left( \frac{v_{\phi,n}}{R} \right)_z =
\frac{v_{\phi,n}}{\chi_n R^2} \hat z,
\label{eq:GradV}
\end{equation}
where $\chi_n \approx 1$. Namely, the distance along the $\hat z$ direction over which $v_{\phi,n}$ changes is $\chi_n R$.

I take the poloidal magnetic field component from equation (\ref{eq:seedN}), and for the gradient of the flow from equation (\ref{eq:GradV}), and substitute both in {{{{ the absolute value of }}}} equation (\ref{eq:Induction2}). This gives the {{{{ magnitude of the }}}} azimuthal magnetic field component at the end of accretion episode $n$
\begin{equation}
 B_{\rm \phi, n} \simeq
 B_{\phi1,n-1} \sin \beta_n   \frac{v_{\phi,n}}{\chi_n R}  \Delta t_n,
\label{eq:Bphi}
\end{equation}
where $\Delta t_n$ is the duration of accretion episode $n$.

I emphasis here that equation (\ref{eq:Bphi}) represents the amplification due only to the change of the angular momentum axis. There are two other effects, which are the usual $\omega$ effect, resulting from the velocity gradient during the considered accretion episode, and the $\alpha$ effect due to turbulence within the accretion flow.

There are two considerations that reduce the effective volume in which the stochastic accretion flow amplifies the magnetic field.
(1) The flow in episode $n$ stretches the seed magnetic fields of episode $n-1$ in the regions where the two toroidal flows cross each other. This is not the entire volume if $\beta_n >0$. (2) The interaction between two consecutive accretion episodes is in the interface between them.
During each accretion episode the regular $\alpha \omega$ dynamo takes place, as mentioned above.

I take the effective volume in which the S$\omega$ effect operates to be a fraction $\delta \ll 1$ of the entire volume of the accretion flow. Only 3D numerical simulations will be able to find the typical value of this parameter. Presently I take it as unknown.

I derive the amplification factor $F_{\rm S \omega}$ due to the S$\omega$ effect during the entire accretion process, from an initial magnetic field $B_{0,0}$ to a final one of $B_{f,{\rm S}\omega}$, by substituting for $N_{\rm ae}$ accretion episodes in equation (\ref{eq:Bphi}), by multiplying by the effective amplification volume fraction $\delta$, and by averaging over the relevant quantities
\begin{equation}
F_{\rm S \omega} \equiv \frac{ B_{f,{\rm S}\omega} }{ B_{0,0} } \approx
   \frac{v_{\phi}}{\chi R} N_{\rm ae}  \Delta t \delta  \sin \beta .
\label{eq:Fsomega}
\end{equation}
In equation (\ref{eq:Fsomega}) the quantities, $v_\phi$, $\sin \beta$, $\chi$, $\delta$, $\Delta t$ are their respective values averaged over the $N_{\rm ae}$ accretion episodes

% ==========================================================
\section{Quantitative estimates}
\label{sec:StochasticOmega}
% ==========================================================

% ===============================
\subsection{Plausible numerical values}
\label{subsec:Values}
% ===============================

I now very crudely estimate the values of the different parameters that appear in equation (\ref{eq:Fsomega}) for the amplification factor of the S$\omega$ effect in the jittering jets explosion mechanism of CCSNe.
I scale equation (\ref{eq:Fsomega}) near the surface of the newly born NS at $R \simeq 20 \km$ as follows
\begin{eqnarray}
\label{eq:ScaledFsomega}
%\begin{aligned}
F_{{\rm S} \omega}  \approx & 20
   \chi^{-1}
   \left(\frac {v_{\phi}} {v_{\rm Kep}} \right)
   \left( \frac {R}{20 \km} \right)^{-3/2}
   \left( \frac {M_{\rm NS}}{1.4 M_\odot} \right)^{1/2}
  \\ &
\times
   \left( \frac{N_{\rm ae}}{10} \right)
   \left( \frac{\Delta t}{0.1 \s} \right)
   \left( \frac{\delta}{0.01} \right)
\left( \frac{\sin \beta}{0.5} \right) .
   \nonumber
%\end{aligned}
\end{eqnarray}
{{{{ The final radius of the NS is about $12 \km$, but during the accretion process the NS is still hot and its radius is somewhat larger than its final radius, hence I scale with $R=20 \km$. }}} }
I elaborate on the scaling of the other different quantities below (see Table \ref{Table1}), and then I compare to some non-dimensional ratios in the solar $\alpha \omega$ dynamo.
\begin{table}[]
\begin{tabular}{|l|c|c|}
\hline
Quantity & Symbol & Crude value \\
\hline
Velocity variation distance  & $\chi R$           & $R\simeq20 \km$ ($\chi \approx 1$)   \\
Toroidal velocity            & $v_\phi$           & $v_\phi \la v_{\rm Kep}$ \\
Number of accretion episodes & $N_{\rm ae}$       &  $10 - 50$ \\
One episode duration         & $\Delta t $        &  $0.03-0.3 \s$ \\
Angle between two episodes   &  $\beta$           &  $30^\circ$ \\
Fraction of effective volume & $\delta$           &  $0.01$ \\
Amplification factor         &$F_{{\rm S} \omega}$&  $ 20 $\\
\hline
\end{tabular}
\caption {The typical parameters of equation (\ref{eq:ScaledFsomega}), which are the typical values averaged over $N_{\rm ae}$ accretion episodes, and the typical amplification factor in the last line (see also Fig. \ref{fig:NSmagneticFig1}).
The typical Keplerian velocity on the surface of the NS is $v_{\rm Kep} \simeq 10^5 \km \s^{-1}$. The relation $N_{\rm ae} \Delta t \approx 1-3 \s$ holds for the entire operation time of the jittering jets. }
\label{Table1}
\end{table}
       
\textit{The distance scale of velocity variation $\chi R$.} I simply assume that the velocity varies along the direction perpendicular to the toroidal flow over a distance of $\simeq R$, i.e., $\chi=1$. This can be smaller even, but then the effective volume fraction $\delta$ might be smaller (see below).  I do note that the velocity gradient between the accreted gas and the surface of the NS might be much larger because over a short radial distance the velocity changes from a slowly rotating NS to $v_\phi$. This, however, is related to the  $\alpha \omega$ dynamo as it does not directly need the stochastic angular momentum accretion. It indirectly requires the stochastic angular momentum accretion to ensure that the newly born NS is a slow rotator and its angular momentum is not aligned with that of the accreted gas.

\textit{The toroidal velocity $v_\phi$.} I scale it with the Keplerian velocity at a radius of $R$ around a newly born NS of mass $M_{\rm NS}$. The velocity in the jittering jets explosion mechanism might be lower than the Keplerian velocity (e.g., \citealt{SchreierSoker2016, Soker2019SASI}), even by a factor of a few. In that case though, the accretion flow is thicker and the effective volume fraction of the S$\omega$ effect $\delta$ will be larger (see below).

\textit{The number of accretion episodes $N_{\rm ae}$ and their average duration $\Delta t$.} The total duration of the explosion process is about a second to few seconds. The number of episodes might be somewhat larger. In that case the average duration $\Delta t$ is smaller, such that $N_{\rm ae} \Delta t \simeq 1-3 \s$.

\textit{The angle between consecutive accretion episodes $\beta$.} The angle is not completely random \citep{PapsihSoker2014Planar} but tends to be smaller than the average value of completely random angular momentum directions. It can be smaller than $\sin \beta =0.5$, but then the overlap between the toroidal flow regions of consecutive episodes is larger, and hence $\delta$ will be larger.

\textit{The fraction of effective volume $\delta$.} The S$\omega$ effect operates when the symmetry axis of the toroidal flow changes direction. We can think of a torus-like region through which there is a toroidal flow of accretion episode $n-1$. This might even be the surface of the newly born NS. Then there is the torus-like region of accretion episode $n$. They each have a volume of ${\rm Vol}_n$. The two torus-like volumes are incline to each other, and hence overlap in a small fraction of the volume $\delta_i {\rm Vol}_n$. In addition the stretching of the magnetic field lines of the torus-like region of episode $n-1$ by the flow in accretion episode $n$ occurs in the interface between them. This is a small fraction $\delta_w $ of the width of the torus. Overall, the S$\omega$ effect operates in a volume that is a fraction of $\delta=\delta _i \delta _w$ of the inflow volume. This value is highly uncertain, and I simply take $\delta \approx 0.01$.

{{{{ The value of $\delta$ cannot be much smaller, as this requires a thin accretion disk at each episode. This can be the case only if the accreted gas has a specific angular momentum that allows it to form an accretion disk. This in turn can be the case only if the angular momentum of the pre-collapse core was high. This brings the situation to another regime of the jet feedback mechanism where there is a more or less constant angular momentum axis. I do not consider this case here (this case will make jet launching easier even). }}}}
The effective volume fraction $\delta$ can be made larger by considering regions with lower toroidal velocity, {{{{ e.g., an accretion belt rather than an accretion disk. }}}} This will reduce $v_\phi$. The value of $\delta$ is larger if on average two consecutive accretion episode are almost aligned with each other. But this will make $\sin \beta$ smaller.

% ===============================
\subsection{Some hints from the solar $\alpha \omega$ dynamo}
\label{subsec:SolarDynamo}
% ===============================

In the S$\omega$ effect the time scale of magnetic field stretching is $t_{\rm B,str} \simeq 2 \pi R/v_\phi $ which is about equal to the Keplerian time $t_{\rm Kep}\simeq 0.001 \s$, or somewhat longer. For the parameters I use here this time scale is $t_{\rm B,str} \approx 0.002 \s$ or somewhat longer. {{{{ I crudely estimate the stretching of the filed during the time $t_{\rm B,str}$ to be by a distance of $2 \pi R \sin \beta$ . }}}}
In the jittering jets explosion mechanism there are several to few tens of accretion episodes over a total time of about a second to several seconds (e.g., \citealt{PapishSoker2011}). Each episode lasts for a time of $\Delta t \simeq {\rm few} \times 0.01 \s $ to $\Delta t \simeq {\rm few} \times 0.1 \s$. The stretching of magnetic field lines between two consecutive accretion episodes lasts for $\approx 0.01-0.1 \s$, which is $\approx 3-30$ times the stretching time $t_{\rm B,str}$ at a radius of $R \simeq 20 \km$.
Taking ten to several tens of accretion episodes, I find that the activity of the S$\omega$ effect lasts for
\begin{equation}
t_{{\rm S} \omega} \approx (30-300) t_{\rm B,str} .
\label{eq:tsomega}
\end{equation}

Let us consider the stretching and entangling timescales in the Sun. In main sequence stars the strength of the magnetic activity is related to the Rossby number ${\rm Ro}$ (or to the dynamo number $N_{\rm D} = {\rm Ro}^{-2}$; e.g., for the Sun, \citealt{KimDemarque1996, Landinetal2010}).
The Rossby number is defined as ${\rm Ro} \equiv P_{\rm rot} / \tau_c$, where $P_{\rm rot}$ is the rotation period and $\tau_c = \alpha_{\rm ml} H_P/v_c \simeq H_P/v_c $ is the local convective turnover time. Here $\alpha_{\rm ml} H_P$ is the mixing length, $H_P$ is the pressure scale height, and $v_c$ is the convective velocity.
The magnetic activity of main sequence stars increases as the Rossby number decreases, until a saturation for ${\rm Ro} \la 0.1$ (e.g., \citealt{Pizzolatoetal2003}).

For solar like stars, the values are $P_{\rm rot} \simeq 25~{\rm day} \approx 200 P_{\rm Kep}$, where $P_{\rm Kep}$ is the Keplerian orbital period on the surface of the star, and $\tau_c \simeq 20~{\rm days}$ (e.g., \citealt{Landinetal2010}).
In the Sun itself the magnetic cycle period is about 22 years (e.g., \citealt{HowardLabonte1980}). Most of the rise in the intensity of magnetic activity within each half a cycle occurs within several years,
$t_{{\rm rise}, \odot} \approx 50 P_{{\rm rot}, \odot} \simeq 60 \tau_{c, \odot}$.

Overall in the Sun, the surface magnetic field intensity rises on a timescale of several tens times the stretching time scale of the field lines. Considering equation (\ref{eq:tsomega}) in relation to this ratio, hints that the total time of operation of the S$\omega$ effect in the jittering jets explosion mechanism allows this mechanism to contribute to the amplification of the magnetic fields in the material that the newly born NS accretes.
This can increase the initial magnetic field of newly born NSs by {{{{ an order of magnitude. The final field intensity depends on other factors beside the operation of the S$\omega$ dynamo. }}}}

% ==========================================================
\section{Discussion and Summary}
\label{sec:Summary}
% ==========================================================

The evolution of magnetic fields from core collapse to NS formation involves four phases of magnetic field amplification.
(1) In the pre-collapse core where a dynamo in the convective zones amplifies magnetic fields (e.g., \citealt{Wheeleretal2015}) and radiative zones store magnetic fields till collapse \citep{Peresetal2019}.
(2) During the collapse itself where the converging inward flow amplifies the radial component of the magnetic fields, {{{{ as magnetic flux conservation implies. }}}}
(3) In the unstable region behind the stalled shock, where in particular the spiral-SASI can amplify the magnetic fields (e.g., \citealt{Endeveetal2010, Endeveetal2012, Rembiaszetal2016a, Rembiaszetal2016b, Obergaulingeretal2018}). (4) Near and on the surface of the newly born NS, e.g., \cite{ObergaulingerAloy2017} who consider only axisymmetrical effects.
In the present paper I addressed the last magnetic field amplification phase.

I considered the contribution of the stochastic angular momentum of the accreted mass to the magnetic field amplification as the mass reaches the surface of the NS. Figure \ref{fig:NSmagneticFig1} presents the basic process, that I term the Stochastic-$\omega$ (S$\omega$) effect.
The toroidal (azimuthal) flow of two consecutive accretion episodes are inclined to each other. As a result of that the toroidal flow of the later episode stretches the magnetic field lines that the early toroidal flow amplified. Within each accretion toroidal flow the regular $\alpha \omega$ dynamo might operate.

Simulations (that do not consider the jittering jets explosion mechanism) find stochastic angular momentum accretion onto newly born NSs (e.g., \citealt{Kazeronietal2016, Mulleretal2017}).
As the jittering jets explosion mechanism must include accretion of stochastic angular momentum with large amplitudes, the S$\omega$ effect is expected to take place in the jittering jets explosion mechanism. I derived an approximate expression for the extra magnetic field amplification of the S$\omega$ effect in equation (\ref{eq:Fsomega}), and substitute typical values (with large uncertainties) in equation (\ref{eq:ScaledFsomega}). This equation suggests that in many cases, {{{{ for which the typical values of the different parameters are crudely listed in Table \ref{Table1}, }}}} the jittering jets explosion mechanism comes along with the formation of a NS with strong magnetic fields. 

The stochastic angular momentum of the accreted gas implies that in many cases the newly born NS will have a slow rotation (relative to breakup rotation velocity).
Overall, according to the jittering jets explosion mechanism many NSs are born with strong magnetic fields, being even magnetars, but with a slow rotation.
In a recent study \cite{Beniaminietal2019} conclude that a fraction of $0.4^{+0.6}_{-0.28}$ of NSs are born as magnetars with magnetic fields at birth of $B \ga 3 \times 10^{13} \G$. They claim that this high fraction challenges existing theories for forming magnetars, as these theories require extreme and rare conditions, {{{{ i.e., pre-collapse rapid rotation and/or strong magnetic fields. }}}} The challenge is stronger even if we take into account that the initial rotation period of most NSs are two orders of magnitudes longer than their maximum possible period (breakup period; e.g., \citealt{PopovTurolla2012, IgoshevPopov2013, Gullonetal2015}).
I here propose that the S$\omega$ effect that operates in the jittering jets explosion mechanism, and even in cases where jets are not launched, might account for the finding of \cite{Beniaminietal2019}  that many NSs are born as magnetars.

% FFFFFFFFFFFFFFFFFFFFFFFFFFFFFFFFFFFFFFFFFFFFFFFFFFFFFFFFFFFFFFFFFFFFFFFFFFF
%% The "ht!" tells LaTeX to put the figure "here" first, at the "top" next
%% and to override the normal way of calculating a float position
%%N%% \begin{figure}[ht!]
%%% One panel accross all page.
%%N%% \plotone{cost-eps-converted-to.pdf}
%%N%% \caption{The subscription and author publication costs from 1991 to 2013.
%%N%% The data comes from Table \ref{tab:table}.\label{fig:general}}
%%N%% \end{figure}
% FFFFFFFFFFFFFFFFFFFFFFFFFFFFFFFFFFFFFFFFFFFFFFFFFFFFFFFFFFFFFFFFFFFFFFFFFFF

% ========================================================================
%%\begin{figure}
% two panels entire page
%%N%% \plottwo{RS_Oph-eps-converted-to.pdf}{U_Sco-eps-converted-to.pdf}
%%N%% \caption{Swift/XRT X-ray light curves of RS Oph and U Sco which represent
%%N%% the two canonical recurrent types, a long period system with a red giant
%%N%% secondary and a short period system with a dwarf/sub-dwarf secondary,
%%N%% respectively.
%%N%% \label{fig:f2}}
%%\end{figure}
%========================================================================

\acknowledgments
I thank Avishai Gilkis for helpful comments,  {{{{ and an anonymous referee for many useful and detailed comments. }}}}
This research was supported by  a grant from the Israel Science Foundation.

%% This command is needed to show the entire author+affilation list when
%% the collaboration and author truncation commands are used.  It has to
%% go at the end of the manuscript.
%\allauthors

%% Include this line if you are using the \added, \replaced, \deleted
%% commands to see a summary list of all changes at the end of the article.
%\listofchanges

\end{document}